\documentclass[preprint,amsmath,noshowkeys,noshowpacs,amssymb,longbibliography]{revtex4-1}

\usepackage{graphicx,color}
\usepackage{cleveref}
\usepackage{amsbsy}
\usepackage{amsthm,mathrsfs,amsopn}
\usepackage{mathtools}
\usepackage{cases}
\usepackage{float}
\usepackage{bbm}

\begin{document}

\title{Stochastic quasi-cycles as a simple explanation for the  time evolution of the Cape Rodney-Okakari Point Marine ecological reserve}

\author{C\'esar Parra-Rojas$^1$, Duccio Fanelli$^2$, Alan J. McKane$^3$ \vspace*{.25cm}}

\affiliation{$^1$ SIRIS Lab, Research Division of SIRIS Academic, 08003 Barcelona, Spain}
\affiliation{$^2$ Dipartimento di Fisica e Astronomia, Universit\`a di Firenze, INFN and CSDC, Via Sansone 1, 50019 Sesto Fiorentino, Firenze, Italy}
\affiliation{$^3$ School of Physics and Astronomy, The University of Manchester, Manchester, M13 9PL, United Kingdom}

\begin{abstract}  The dataset collected at the Cape Rodney-Okakari Point Marine (CR-OPM) reserve on the North Island of New Zealand is rather unique. It describes the cyclic time evolution of a rocky intertidal community, with the relative abundances of the various coastal species that have been meticulously monitored for more than 20 years. Past theoretical studies, anchored on a deterministic description, required invoking ad hoc mechanisms to reproduce the observed dynamical paths. Following  a maximum likelihood approach to interpolate individual stochastic trajectories, we here propose quasi-cycles as an alternative and simpler mechanism to explain the oscillations observed in the population numbers of the ecosystem.  From a general standpoint, we also show that it is possible to return conclusive evidence on the existence of stochastic quasi-cycles, without resorting to global fitting strategies which necessitate handling a large collection of independent replicas of the  dynamics, a possibility that is often precluded in real life applications. \end{abstract}

\maketitle

\section{Introduction}
\label{sec:intro}

Studying the interlaced dynamics of mutually entangled species defines a field of paramount importance and cross disciplinary interest. Accurate experimental data collected in the field are usually guarded as a precious asset, with the aim of providing solid benchmarks for model analysis. The temporal scales involved are typical of the system under investigation. To identify recurrent patterns and archetypal features, it might prove necessary to collect data for years---as it is often the case in ecology---thus making the overall process prohibitive. Very distinctive, in this respect, is the dataset collected at the Cape Rodney-Okakari Point Marine (CR-OPM) reserve on the North Island of New Zealand. It describes the cyclic evolution of a rocky intertidal community: the relative abundances of the various coastal species have been constantly acquired for more than 20 years, a meticulous work that resulted in a rather unique database \cite{Ballantine2008, Ballantine2014}.  Three sessile species can be identified: the honeycomb barnacle {\it Chamaesipho columna}, the crustose brown alga {\it Ralfsia cf confusa}  and the little black mussel {\it Xenostrobus pulex}. Barnacles colonize bare rock by gregarious settlement and yield a carpet that covers the rocky surface. Crustose algae settle on top of the barnacles, without harming them. On the contrary, Xenostrobus mussel larvae settle gregariously on top of barnacles and crustose algae. In doing so they develop a compact sheet, that kills the barnacles underneath. The dead barnacles detach from the rock: as the mussel carpet is no longer anchored to a solid surface, it is washed away by the tides. Hence, bare rock becomes available again, for the next cycle in the succession. The species dynamics has been thoroughly analyzed to understand the displayed cyclicity,  following the scheme outlined above. Noteworthy is the work by Beninc{\`a} et al. \cite{beninca2015} where a patch-occupancy model with seasonal forcing was developed to claim a field demonstration of non-equilibrium coexistence of competing species through a cyclic succession at the edge of chaos.

 The classical approach to population dynamics, including the aforementioned work by  Beninc{\`a} et al \cite{beninca2015}, relies however on quantitatively characterizing the densities of the involved species through a system of ordinary differential equations. As opposed to this formulation, a different level of modeling can be invoked which favors an individual-based description, and is thus intrinsically stochastic ~\cite{vankampen2011, gardiner2009}. The microscopic dynamics is characterized in terms of the so-called master equation, a balance update rule that returns the probability for observing the system in a given state, at a given time. Master equations can only be solved exactly in very few cases; different approximation schemes have been therefore devised to elaborate the subtle role played by demographic noise. Endogenous stochasticity can for instance amplify via a self–consistent mechanism, driving the emergence of almost regular oscillations, termed in the literature quasi-cycles~\cite{PhysRevLett.94.218102, PhysRevE.96.022308, PhysRevE.96.062313, PhysRevE.90.032135, PhysRevE.84.051919, PhysRevE.79.036112}.  Noise can hence organize in regular quasi-periodic orbits, building macroscopic order from microscopic disorder. Starting from these premises, we will here propose and test an alternative mathematical framework to account for the dynamics of the species that populate the CR-OPM reserve. The proposed interpretative scenario takes a stochastic standpoint, as opposed to other models discussed in the literature.

From a general perspective, and once the stochastic model has been established, one can operate  under the linear noise approximation. This ultimately amounts to considering moderate stochastic effects to analytically access the power spectrum of fluctuations. The latter can be compared to the corresponding quantity as determined experimentally so as to obtain a direct estimate of the parameters involved.  This is a global fitting strategy which requires ensemble averages of independent experimental outputs to be carried out. Another strategy, that will be utilized here, aims at fitting individual stochastic trajectories~\cite{zimmer2015a, zimmer2015b}, as recorded experimentally. This approach is to be favored when independent replicas of the  stochastic system cannot be accessed, as is the case for the CR-OPM reserve database. In more detail, the core idea is to implement a maximum likelihood approach that seeks to interpolate the multi-dimensional distribution of fluctuations with a multivariate Gaussian, as obtained under the linear noise approximation \cite{zimmer2015a, zimmer2015b}. This procedure returns a best fit interpolation for the individual stochastic time series with an associated determination of the parameters involved, including the (otherwise elusive) system size. Working along these lines, we will show that the Cape Rodney-Okakari Point Marine Reserve data are well fitted by a basic stochastic model, with no inclusion of external forcing. This model provides a simpler, though equally accurate, explanation of the data. As such, it should be preferred over other available options by virtue of a general principle of parsimony. More  generally, the results here discussed report on the first attempt to perform a stochastic interpolation of an experimental time series, whose seemingly regular oscillations are attributed to the resonant amplification of noisy quasi-cycles.

\section{Methods}
\subsection{Generic setting: Model formulation}

In the following we will begin by considering a generic, microscopic-level representation of the dynamics. Our notation and methodology follows Refs.~\cite{mckane2014,BLACK2012337}. 

Interacting individuals are organized in one or more different subpopulations, which we describe by a state-space variable, $\bm n$, where $n_i$ is the number of individuals in subpopulation $i$. The interactions between individuals in the system, each of which induces a change in their numbers, can be modeled as a series of chemical-like reactions with their associated transition rates; that is, the $\mu$-th reaction occurs at a rate $T_\mu$ and produces a change in the population numbers given by the stoichiometric vector $\bm \nu_\mu$. The fluctuations in the system thus defined stem directly from the microscopic interactions between individuals, and are a consequence of its finite size $N$. Here, $N$ quantifies the total number of individuals belonging to the system under consideration (or a related quantity which scales extensively with the volume of the system). The time evolution of the occupation levels of each subpopulation can be expressed probabilistically as a master equation~\cite{vankampen2011}; however, a more tractable approach corresponds to describing the time evolution of the system in terms of population \emph{densities}, as opposed to absolute numbers. This entails approximating the discrete state-space defined by $\bm n$ by a continuous state-space, given by $\bm x\equiv \bm n/N$, for large-but-finite system size $N$. This mesoscopic-level representation can be described by the following stochastic differential equation (SDE)~\cite{gardiner2009}:
\begin{align}
\begin{split}
\dot{\bm x} &= \bm A(\bm x) + \frac{1}{\sqrt{N}}\bm \eta\\
\left\langle \eta_i(t) \eta_j(t')\right\rangle &= B_{ij}(\bm x)\delta(t-t'),
\end{split}\label{eqn:sde}
\end{align}
where the dot represents the time derivative, and
\begin{align}
    \begin{split}
            \bm A(\bm x) &= \sum_{\mu}\bm \nu_{\mu}f_{\mu}(\bm x)\\
            \bm B(\bm x) &= \sum_{\mu}\bm \nu_{\mu}\bm \nu_{\mu}^\top f_{\mu}(\bm x),
    \end{split}\label{eqn:Ai}
\end{align}
with $f_{\mu}(\bm x)$ identifies the transition rate of process $\mu$,  with $\bm n/N$ replaced by $\bm x$~\cite{mckane2014}. Here $\bm \eta$ is Gaussian white noise with zero mean.

The deterministic solution to the equation above is obtained by performing the so called mean field (or, equivalently, thermodynamic) limit. This amounts to requiring $N\to \infty$ which readily yields $\dot{\bm x}_{\rm det} = \bm A(\bm x_{\rm det})$, where the label attached to the state variable has been inserted so as to recall that  under this condition one operates in a deterministic setting. By linearizing the system around the deterministic trajectory for large sizes $N$, we find that the stochastic perturbations $\bm \xi(t) \equiv \sqrt{N}(\bm x(t)-\bm x_{\det}(t))$, satisfy~\cite{mckane2014}
\begin{align}
\begin{split}
    \dot{\bm \xi} &= \bm J(\bm x_{\rm det})\cdot \bm \xi + \bm{\bar{\eta}}\\
    \left\langle \bar{\eta}_i(t) \bar{\eta}_j(t')\right\rangle &= B_{ij}(\bm x_{\rm det})\delta(t-t')
\end{split}\label{eqn:lna}
\end{align}
to first order in $1/\sqrt{N}$, where $\left\langle \bm{\bar{\eta}}\right\rangle = \bm 0$ and $\bm J(\cdot)$ is the Jacobian of $\bm A$. The mean and covariance matrix of these perturbations follow
\begin{align}
\begin{split}
    \dot{\left\langle \bm \xi \right\rangle} &= J(\bm x_{\rm det})\cdot \left\langle \bm \xi\right\rangle\\
    \dot{\bm \Xi} &= \bm J(\bm x_{\rm det})\cdot \bm \Xi + \bm \Xi \cdot \bm J(\bm x_{\rm det})^\top + \bm B(\bm x_{\rm det}),
\end{split}\label{eqn:lna2}
\end{align}
with $\bm \Xi(t) \equiv \left\langle (\bm \xi(t)-\left\langle \bm \xi \right\rangle (t))(\bm \xi(t)-\left\langle \bm \xi \right\rangle (t))^\top \right\rangle$. This is the linear noise approximation (LNA), which we will use in the following sections to estimate model parameters and reconstruct stochastic trajectories from data.

\subsubsection{An illustrative example: The Lotka--Volterra model}

As an illustration of the above machinery, we select a simple Lotka--Volterra model describing predator-prey dynamics (see Appendices for details about the mathematical setting). In this case, the state vector is given by $\bm x\equiv (x,y)$, where $x$ and $y$ represent the population densities of prey and predator, respectively. We assume that the predator population dies at unit rate, and that it converts food into offspring with 100\% efficiency, so that the only parameters governing the dynamics are the prey's growth rate and the rate of predation, which we denote by $\beta$ and $\gamma$, respectively. With this, the system satisfies the SDE~\eqref{eqn:sde} with (see SI)
\begin{align}
\begin{split}
    A_x(\bm x) &= \beta x(1-x-y)-\gamma x y\\
    A_y(\bm x) &= \gamma x y - y\\
    B_{xx}(\bm x) &= \beta x(1-x-y)+\gamma x y\\
    B_{xy}(\bm x) = B_{yx}(\bm x) &= -\gamma x y\\
    B_{yy}(\bm x) &= \gamma x y + y.
\end{split}\label{eqn:lv}
\end{align}
The deterministic limit of this system displays a fixed point at $(x,y)=\gamma^{-1}(1, \beta(\gamma - 1)(\beta+\gamma)^{-1})$, which is approached via damped oscillations for $\gamma > (1 + \sqrt{1 + \beta})/2$. Therefore, we expect quasi-cycles to occur in the stochastic system for a choice of parameters values within this region. Quasi-cycles are seemingly 
regular oscillations for the densities of the interacting species, as  triggered by the noisy component of the dynamics. More specifically, quasi-cycles are a manifestation of a resonant amplification mechanism that enhances the endogenous fluctuations as stemming from finite-size corrections to the underlying deterministic dynamics.

\begin{figure*}
    \centering
    \includegraphics[width=0.8\textwidth]{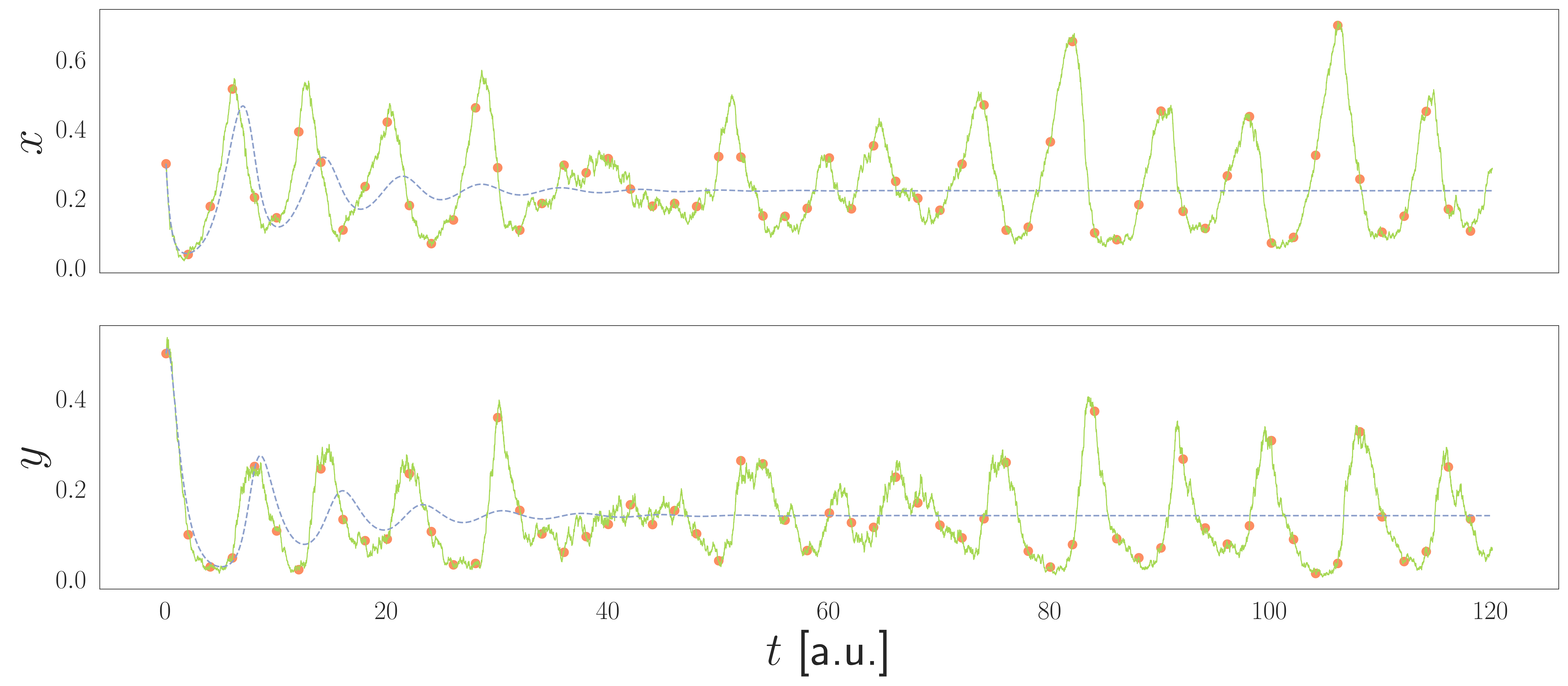}
    \caption{Stochastic realization from the Lotka--Volterra model~\eqref{eqn:lv} (continuous line), with its corresponding deterministic trajectory (dashed line) and sampled data (circles). Parameter values: $\beta = 1$, $\gamma = 4.5$, and $N=200$.}
    \label{fig:toy_data}
\end{figure*}

Stochastic trajectories for the model under consideration are generated in three different ways: direct simulation of the microscopic system with the Gillespie algorithm~\cite{gillespie1977}; simulation of its equivalent SDE---see Fig.~\ref{fig:toy_data}; and simulation of the fluctuations around the fixed point of the system with the LNA.  The results we obtain in the following sections are equivalent, regardless of the generating mechanism, provided $N$ is sufficiently large for the linear noise approximation to hold. 

The stochastic trajectory, as obtained via direct simulations, can be further sampled to artificially re-create the discreteness of data measured in a real system, see Fig.~\ref{fig:toy_data}. This discrete trajectory constitutes the `artificial data' used as an input of the analysis carried out in the following. In  Fig.~\ref{fig:toy_data} the obtained stochastic trajectories are compared to the corresponding deterministic curves as obtained in the infinite $N$ limit. As it can clearly appreciated by visual inspection, the deterministic curves converge to the  fixed points 
 via damped oscillations, as opposed to their stochastic counterparts that display self-sustained, though irregular, oscillations.

\subsection{Parameter estimation}\label{sec:estimation}

We will start from a dataset consisting of a total of $K$ measurements of population counts, taken at different times $t_i$, with $i=0,1,\ldots,K-1$, which need not be equidistant. Let us denote by $\bm \rho_i$ the vector of coordinates obtained at time $t_i$. Further  $\bm \theta$ stands for the vector which encapsulates the free parameters of the stochastic model in~\eqref{eqn:sde}. In the case of the Lotka--Volterra model described above, we have $\bm \rho_i = (x(t_i), y(t_i))$, and $\bm \theta = (\beta, \gamma)$.

We are interested in finding $\bm{\hat{\theta}}$ which we take to be the set of parameter values that provide the best fit between the model and the collected data. In order to do so, we use the LNA with a \emph{multiple-shooting} approach---see~\cite{zimmer2015a, zimmer2015b, buckingham2018}: for each timepoint $t_i$, $i=1,\ldots,K-1$, we define a likelihood function for the vector of parameters $\bm \theta$, denoted by $L_i(\bm \theta)$, corresponding to the probability of obtaining the current coordinates $\bm \rho_{i}$ starting from an initial condition given by the previous measurement $\bm \rho_{i-1}$ at time $t_{i-1}$ for the respective value of $\bm \theta$; that is,
\begin{align}
    L_{i}(\bm \theta \vert \bm \rho)  &= p(\bm \rho_i \vert \bm \rho_{i-1}, \bm \theta).
\end{align}
Using the LNA, the solution for the state of the system linearized around the deterministic trajectory that starts from the point $\bm \rho_{i-1}$ is obtained through~\eqref{eqn:lna2} with initial condition $\bm \xi = \bm 0$. This means that, at time $t_i$, the probability $p(\bm \rho_i \vert \bm \rho_{i-1}, \bm \theta)$ of finding the system in the state $\bm \rho_i$ given the starting point $\bm \rho_{i-1}$ and the parameters $\bm \theta$ ---in other words, the likelihood $L_i(\bm \theta \vert \bm \rho)$---follows a multivariate Gaussian distribution with mean $\bm \mu_i$ and covariance matrix $\bm \Sigma_i$ given by
\begin{align}
\bm \mu_i &= \bm x_{\rm det}(\Delta t_i; \bm \rho_{i-1}, \bm \theta)\\
\bm \Sigma_i &= N^{-1}\bm \Xi(\Delta t_i; \bm \rho_{i-1}, \bm \theta),
\end{align}
where $\Delta t_i \equiv t_i-t_{i-1}$. The combined likelihood for the parameter values given all the datapoints $\bm \rho_i$, $i=0,1,\ldots,K-1$, will then correspond to the probability of finding the state $\bm\rho_1$ starting from $\bm\rho_0$, and $\bm\rho_2$ starting from $\bm\rho_1$, and so on, for the given parameters, \emph{i.e.},
\begin{align}
    L(\bm \theta \vert \bm \rho) &= \prod _{i=1}^{K-1} L_i(\bm \theta \vert \bm \rho),\label{eqn:likelihood}
\end{align}
and the best-fit parameters will be given by $\bm{\hat{\theta}}~=~\mathrm{arg\,max}_{\bm \theta}~L(\bm \theta \vert \bm \rho)$.

We implement the likelihood in~\eqref{eqn:likelihood} in Python and find its global maximum using the differential evolution algorithm~\cite{storn1997} made available in the SciPy library~\cite{scipy}. For details of the implementation, see \href{https://github.com/cparrarojas/sde-parameter-estimation}{github.com/cparrarojas/sde-parameter-estimation}. 

\subsection{Reconstruction of trajectories}

Once the best-fit parameters $\bm{\hat{\theta}}$ have been obtained, we again use the LNA with multiple shooting in order to illustrate the trajectory followed by the real data as a possible realisation of the underlying stochastic process; this means that we do not fit a line through the datapoints, but rather a \emph{band} representing the fluctuations inherently present in the system. To this end, at each timepoint $t_i$ we generate a distribution of candidate state vectors $\bm x_i$ using \eqref{eqn:lna2} with initial condition $\bm\xi = \bm 0$; that is, we follow the deterministic trajectory of the model starting from the observed data at time $t_{i-1}$, $\bm x_{i-1}=\bm \rho_{i-1}$, and generate a distribution for the fluctuations around the deterministic state at time $t_i$. This is repeated for all $i=1,\ldots,K-1$.

\section{The Cape Rodney-Okakari Point Marine Reserve}

To test the adequacy of the proposed fitting scheme against real data we will consider the dynamics of a rocky intertidal community in the Cape Rodney-Okakari Point Marine Reserve on the North Island of New Zealand. The reserve was established in 1975 and offers ideal conditions for long-term studies of species interactions because of the reduced anthropogenic impacts. As mentioned above, the ecosystem is composed of three sessile species: the honeycomb barnacle \textit{Chamaesipho columna}, the crustose brown alga \textit{Ralfsia cf confusa}, and the little black mussel \textit{Xenostrobus pulex}. Species fluctuations were monitored on a monthly basis for more than 20 years, a painstaking enterprise which resulted in a rich and rather unique dataset for challenging theoretical models. The scheme of interactions among species has been thoroughly characterized, as mentioned earlier. In the following, for clarity,  we recall the main steps involved. Barnacles colonize bare rock, generating extensive sheets that cover the rocky surface. Crustose algae settle on top of the barnacles and initiate a symbiotic relationship, without harming the host.  Xenostrobus mussel larvae cannot attach on smooth bare rock. They can however settle gregariously on top of barnacles and crustose algae. This results in a dense carpet of mussels which eventually kill the barnacles underneath. When the dead barnacles detach from the rock, the mussel carpet lack a solid anchoring  and it is therefore washed away by daily tides. Hence, bare rock is again available and the whole process starts anew. In \cite{beninca2015}, fluctuations in the species abundances, as stemming from  cyclic successions, have been analyzed in the context of a deterministic model. On the basis of data analysis, Beninc{\`a} et al. \cite{beninca2015} suggest in particular that the observed fluctuations originate from the repeated occurrence of stabilizing and chaotic dynamics during species replacement and the findings are interpreted as a field demonstration of nonequilibrium coexistence between competing species at the edge of chaos. Building on the method developed in the early part of the paper, we here aim at testing an interpretative framework which is alternative to that proposed by Beninc{\`a} and collaborators. To this end, we shall consider a stochastic version of the model introduced in \cite{beninca2015} and allow for endogenous, finite-size corrections to the idealized mean-field deterministic picture. As we shall show in the following, the time series recorded at the Cape Rodney-Okakari Point Marine Reserve can be well fitted, via the procedure introduced above, by accommodating for suitable stochastic contributions. In the following we will in particular denote by $B_0$ individuals belonging to the family of barnacles. A barnacle covered by algae is labeled $B_A$. Algae and mussels are respectively denoted by $A_0$ and $M$, while for bare rock the symbol $R$ is used. The size of the system is specified by $N$. The number of individuals belonging to different families, at a given time, are respectively given by  $n_{B_0}$, $n_{B_A}$, $n_{A_0}$, $n_{M}$, $n_{R}$, with an obvious meaning of the adopted notation. The state of the system is consequently represented by the vector $\bm n = (n_{B_0}, n_{B_A}, n_{A_0}, n_{M}, n_{R})$. Further, we  define the associated continuous variables, as $b_0 = n_{B_0}/N$, $b_A = n_{B_A}/N$, $a_0 = n_{A_0}/N$, $m = n_{M}/N$ and $r = n_{R}/N$, or, in compact notation, $\bm x = (b_0, b_A, a_0, m, r)$.

\subsection{Description of the model}

We follow~\cite{beninca2015} and describe the dynamics of the system in terms of the following events: barnacles take over the bare rock at rate $C_{br}$; algae colonize the barnacles at rate $C_{ab}$; algae-covered barnacles, in turn, may also colonize adjacent patches of bare rock at rate $C_{br}$; mussels can invade a patch occupied either by bare or algae-covered barnacles at rate $C_{m}$; algae can colonize bare rock at rate $C_{ar}$; barnacles, algae and mussels die at rates $m_b$, $m_a$ and $m_m$, while they may also invade the system from outside at rates $\alpha_b$, $\alpha_a$ and $\alpha_m$, respectively. The full details of the individual interactions involved in each of these events, plus their stoichiometric vectors, are given in the Appendix, Eqs.~[1]--[20]. It is important to mention that, unlike~\cite{beninca2015}, we do not consider a forcing term acting on the mortality of the mussel population, since the degree of stochasticity inherent to the system alone is able to generate the patterns observed in the data.

We note that, since $n_{B_0}+n_{B_A}+n_{A_0}+n_{M}+n_{R} = N$, we can reduce the problem to a four-dimensional system with a state vector $\bm{n}=(n_{B_0}, n_{B_A}, n_{A_0}, n_{M})$, and $n_{R}=N-n_{B_0}-n_{B_A}-n_{A_0}-n_{M}$. Using \cref{eqn:Ai},
we find that the system defined microscopically in the Appendix satisfies the SDE~\eqref{eqn:sde}, with
\begin{align}\begin{split}
A_{{b}_0} &= C_{br}(b_0+b_A)r + \alpha_b r - C_{ba}(b_A + a_0)b_0 - \alpha_a b_0\\
&\qquad - (C_m m + \alpha_m)b_0-m_bb_0 + m_ab_A\\
A_{{b}_A} &= C_{ba}(b_A + a_0)b_0 + \alpha_a b_0 - (C_m m + \alpha_m)b_A-m_ab_A - m_bb_A\\
A_{{a}_0} &= C_{ar}(b_A + a_0)r + \alpha_a r - (C_m m + \alpha_m)a_0-m_aa_0 + m_bb_A\\
A_{m} &= (C_m m + \alpha_m)(b_0+b_A+a_0)-m_m m,
\end{split}\label{eqn:Ai_beninca}
\end{align}
where $r\equiv 1-b_0-b_A-a_0-m$; the noise correlator is given by
\begin{align}\begin{split}
B_{b_0 b_0} &= C_{br}(b_0+b_A)r + \alpha_b r + C_{ba}(b_A + a_0)b_0 + \alpha_a b_0\\
&\qquad + (C_m m + \alpha_m)b_0 + m_bb_0 + m_ab_A\\
B_{b_0 b_A} &= -C_{ba}(b_A + a_0)b_0 - \alpha_a b_0 -m_ab_A\\
B_{b_0 a_0} &= 0\\
B_{b_0 m} &= - (C_m m + \alpha_m)b_0\\
%B_{b_0 r} &= -C_{br}(b_0+b_A)r - \alpha_b r - m_bb_0\\
B_{b_Ab_A} &= C_{ba}(b_A + a_0)b_0 + \alpha_a b_0 + (C_m m + \alpha_m)b_A+m_ab_A\\
&\qquad  + m_bb_A\\
B_{b_Aa_0} &= - m_bb_A\\
B_{b_Am} &= - (C_m m + \alpha_m)b_A\\
%B_{b_Ar} &= 0\\
B_{a_0a_0} &= C_{ar}(b_A + a_0)r + \alpha_a r + (C_m m + \alpha_m)a_0+m_aa_0\\
&\qquad  + m_bb_A\\
B_{a_0m} &= -(C_m m + \alpha_m)a_0\\
%B_{a_0r} &= -C_{ar}(b_A + a_0)r - \alpha_a r - m_aa_0\\
B_{mm} &= (C_m m + \alpha_m)(b_0+b_A+a_0)+m_m m.
\end{split}\label{eqn:Bij_beninca}
\end{align}

Starting from these premises, we have performed the parameter estimation procedure as outlined in \cref{sec:estimation} for this stochastic, unforced model, using a subset of the CR-OPM reserve dataset. These latter have been pre-processed in order to ensure that the population densities for all species involved are correctly normalized and satisfy the model's constraint $b_0+b_A+a_0+m+r=1$---see Appendix.

\section{Results and Discussion}

To date, stochastic quasi-cycles have been invoked to yield a viable explanation for a limited selection of real life applications, as e.g. the emergence of circadian clocks in cyanobacteria \cite{stavans} (see also \cite{goldenfeld, lavacchi, dipatti} for a spatially extended version of noise seeded instability). In \cite{stavans} the power spectrum of fluctuations as collected from different replicas of the experimental time series was compared with the corresponding theoretical profile, as obtained from the governing master equation, under the linear noise approximation. It should be however remarked that it is not in general straightforward to access a sufficiently large collection of independent experimental realizations of the dynamics. Scarce statistics can indeed limit the effectiveness of a fitting strategy targeted to the underlying power spectrum. To overcome this inherent limitation, while  still addressing the possible existence of quasi cycles in nature, we aim here at  recovering robust estimates of the relevant---\textit{a priori} unknown---kinetic parameters by just interpolating one individual trajectory, using the method discussed above. 

To prove the adequacy of the proposed methodology, we begin by testing it using the predator-prey model, as introduced earlier. This latter model has been  chosen for pedagogical reasons, as a prototypical example of a system that displays the spontaneous emergence of stochastic quasi-cycles. Mock trajectories are generated in silico for a given choice of the reference parameters and sampled at different rates, so as to make contact with different experimental scenarios. More specifically, we estimate the parameters of the model ($\beta$, $\gamma$, and the system size $N$) for the trajectory  sampled at 3 different time intervals---10, 100 and 200 timesteps---in order to obtain 3 differing artificial datasets. As expected,  more confident estimates are obtained for \emph{better quality data}; that is, for data that has been sampled more frequently from the underlying population dynamics, the distributions of the values estimated for the free parameters of the system tend to be more tightly clustered around a central peak near their nominal values.  \cref{fig:toy_estimation} illustrates the results of the fitting procedure for the case of sampling every 100 timesteps, while \cref{fig:toy_reconstruction} shows the \emph{fitted curve}, or cloud of fluctuations, through the data. The cases corresponding to the other sampling rates can be seen in Appendix, Figs.~S1--S2.

\begin{figure*}
    \centering
    \includegraphics[width=0.33\textwidth]{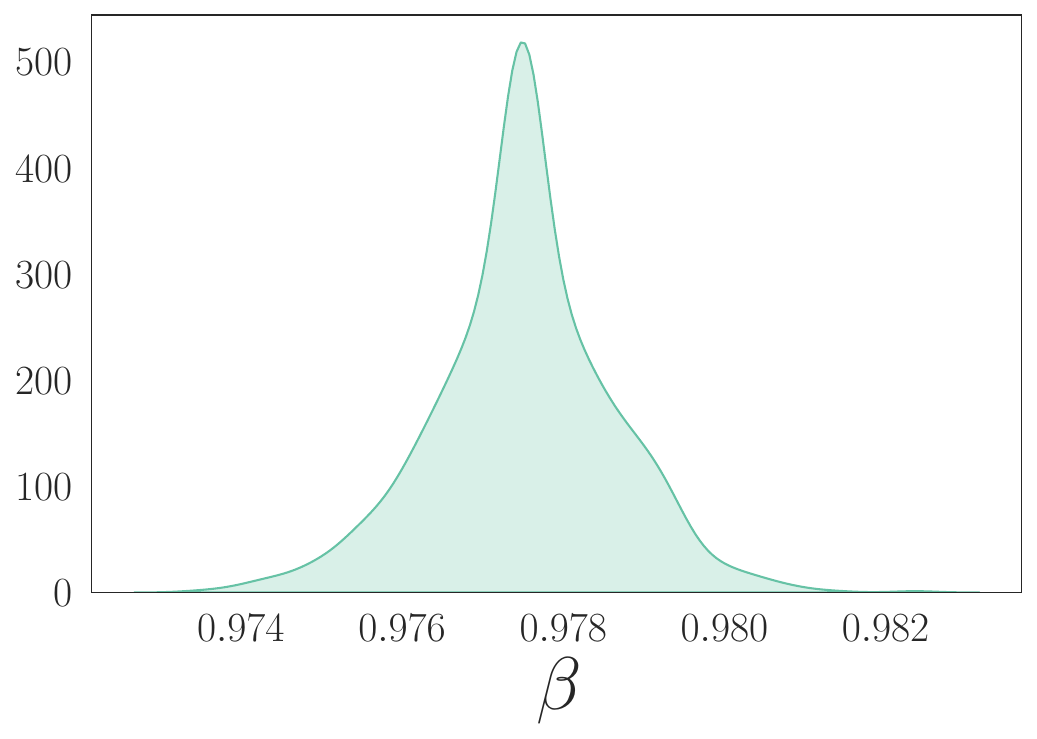}\hfill \includegraphics[width=0.33\textwidth]{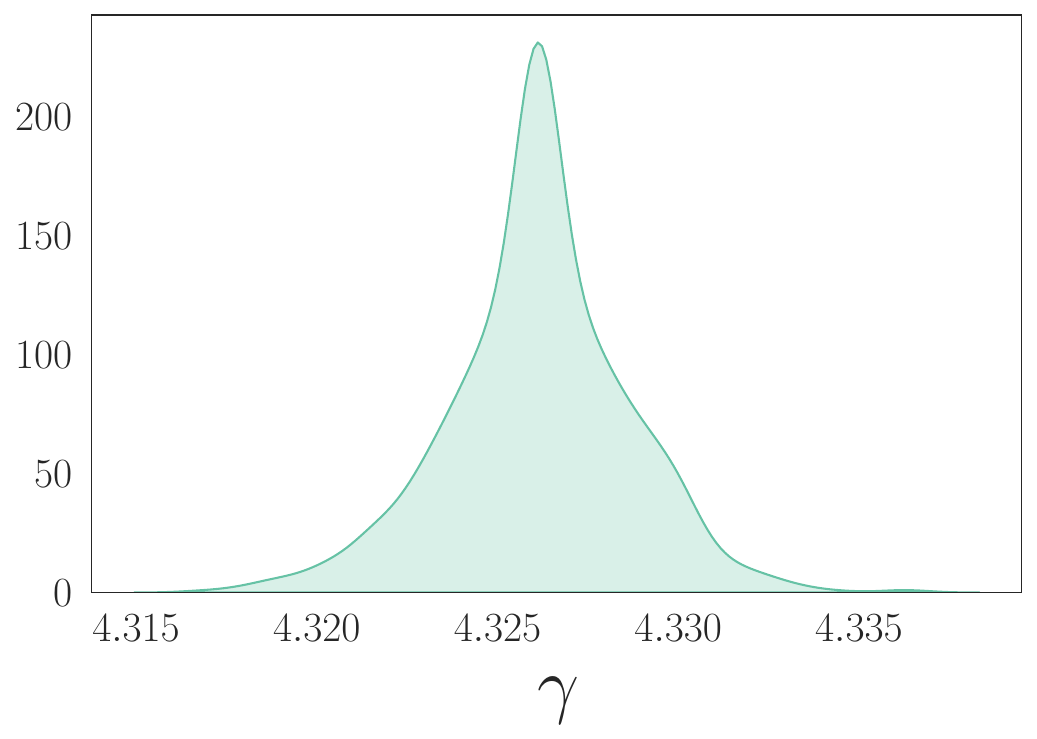}\hfill \includegraphics[width=0.33\textwidth]{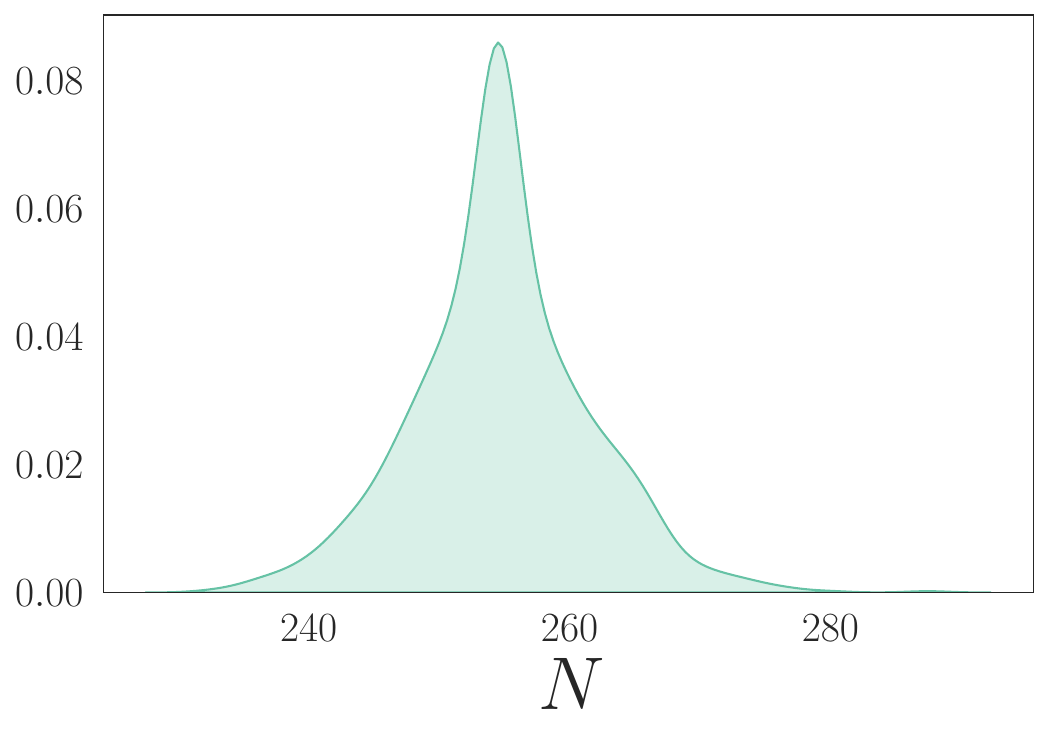}
    \caption{Distribution of the parameters obtained for the data from the Lotka--Volterra model, after estimating it 1200 times with a sampling interval of 100 timesteps. Nominal values: $\beta=1$, $\gamma=4.5$, and $N=200$.}
    \label{fig:toy_estimation}
\end{figure*}

\begin{figure*}
    \centering
    \includegraphics[width=0.9\textwidth]{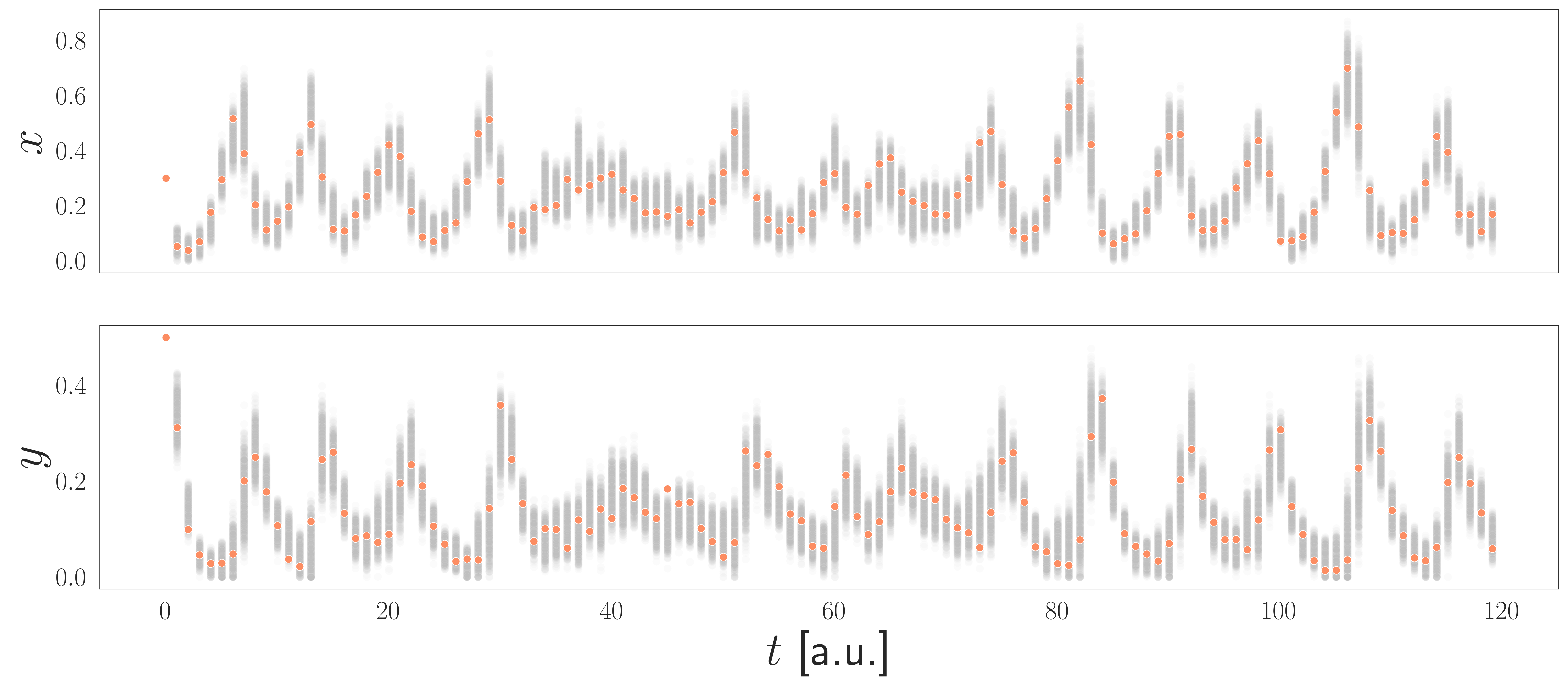}
    \caption{Reconstruction of the Lotka--Volterra time series by using the parameters obtained from the estimation with a sampling interval of 100 timesteps, and 1500 points for the cloud of fluctuations.}
    \label{fig:toy_reconstruction}
\end{figure*}

Motivated by this success against synthetically generated data, we moved forward to considering the single---and rather unique---trajectory that is available for the Cape Rodney-Okakari Point Marine Reserve.

Using likelihood profiles~\cite{raue2009} with a deterministic version of the model, we find that the large number of free parameters of the system precludes a satisfactory fit; due to this, we fix 4 of them: the 3 immigration terms, set to the same small value as in~\cite{beninca2015}. Further, the colonization rate of bare rock by algae, $C_{ar}$ is set to zero because of its tendency to jump towards the lower bound of the estimation range. With these assumptions, we are able to obtain consistent, Gaussian-like estimates for all the other parameters in the model for the actual reserve data. The results are summarised in \cref{tab:data_estimation}.

\begin{table}[]
    \centering
    \begin{tabular}{c|c|c}
         \ & mean & std \\
         \hline 
        $N$ & 15.5 & 1.4 \\
        $C_{br}$ & $7.3\times 10^{-3}$ & $4.7\times 10^{-4}$ \\
        $C_{ba}$ & $2.2\times 10^{-2}$ & $1.6\times 10^{-3}$ \\
        $C_m$ & $6.2\times 10^{-3}$ & $3.8\times 10^{-4}$ \\
        $m_b$ & $6.5\times 10^{-6}$ & $6.2\times 10^{-7}$ \\
        $m_a$ & $5.1\times 10^{-3}$ & $4.9\times 10^{-4}$ \\
        $m_m$ & $2.1\times 10^{-3}$ & $2.0\times 10^{-4}$
    \end{tabular}
    \caption{Mean and standard deviation of the parameter values obtained for the processed CR-OPM data, shown in the Appendix, Fig.~S3, after simultaneously estimating them 1200 times.}
    \label{tab:data_estimation}
\end{table}

\begin{figure}
    \centering
    \includegraphics[width=0.9\textwidth]{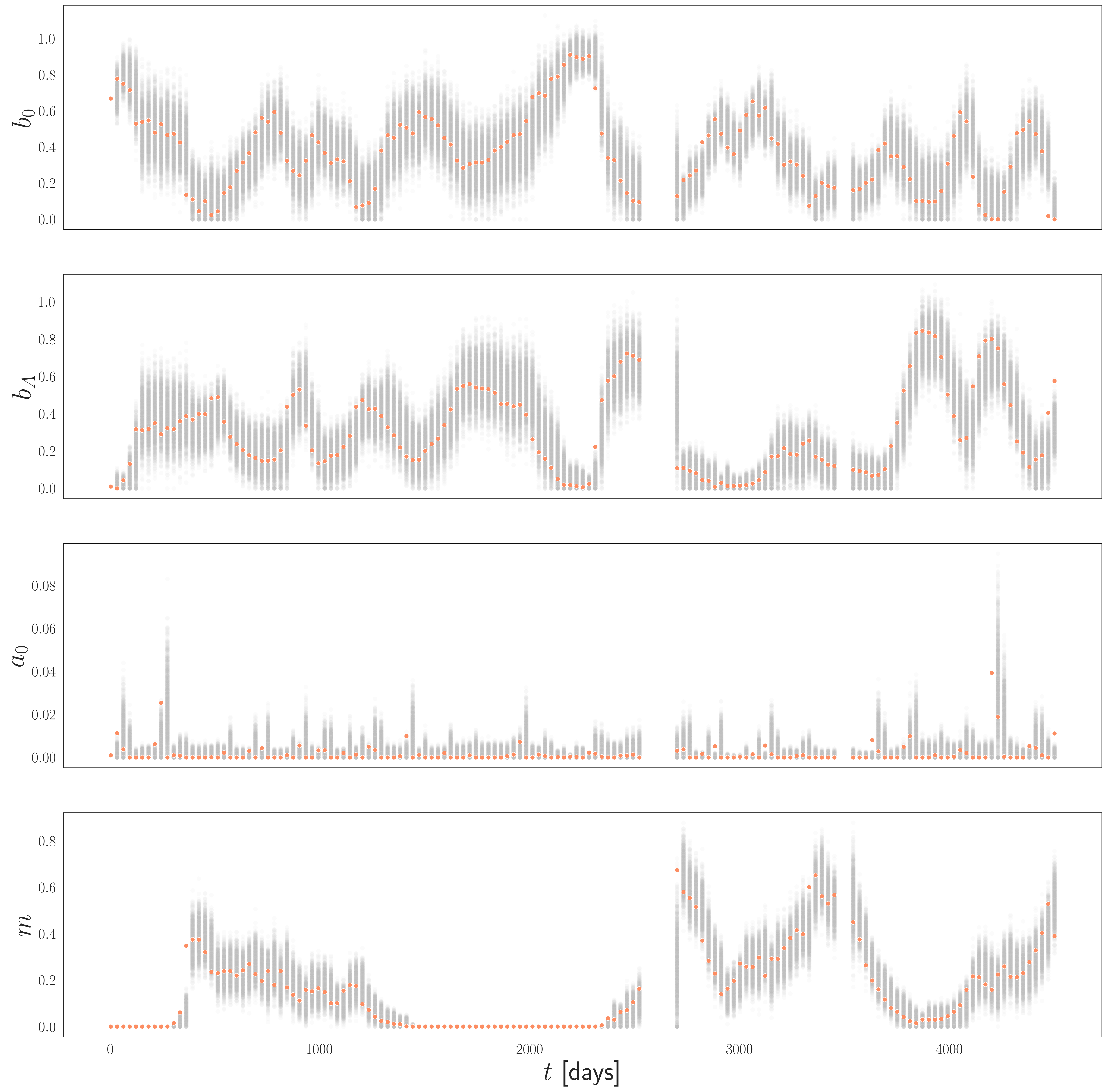}
    \caption{Reconstruction of the time series obtained from the processed CR-OPM data, shown in the Appendix, Fig.~S3, using the means of the estimated parameters in \cref{tab:data_estimation} and 1500 points for the cloud of fluctuations.}
    \label{fig:data_reconstruction}
\end{figure}

We note that the values obtained fall within the region of the parameter space that yields a stable fixed point that is approached via damped oscillations in the deterministic limit. In other words, the best-fit parameters support a scenario in which the recurring oscillations observed in the populations numbers in the reserve are caused by stochastic resonant amplification, without the need for external forcing. The fitted trajectories for all species are shown in \cref{fig:data_reconstruction}.

\newpage

\newpage

Summing up, we have here tested and applied a maximum likelihood approach to interpolate individual stochastic trajectories, as stemming from the  dynamics under investigation. This enables one to extract sensible estimates of the underlying kinetic parameters, in terms of the description of the emergence of quasi-cycles from an endogenous stochasticity drive, without resorting to global fitting strategies.
The proposed method interpolates the multi-dimensional distribution of fluctuations with a multivariate Gaussian, as computed using the  linear noise approximation. Calibration tests are performed against synthetic data generated from a simple stochastic predator-prey model that displays quasi-cycles. The numerical scheme is indeed capable of reproducing the sought-after parameters with an excellent degree of accuracy. We then turned to applying the algorithm to the Cape Rodney-Okakari Point Marine Reserve database. For obvious reasons, independent replicas of the measurements cannot be accessed and thus ensemble fitting strategies do not represent a viable option for this specific case study. Following the strategy outlined above, we were able to show that the Cape Rodney-Okakari Point Marine Reserve data can be well fitted by a basic stochastic model, with no inclusion of external forcing, as was invoked in the context of a deterministic description. Based on the Occam's razor principle, the stochastic interpretation of the available data should be preferred against its deterministic analog, since it returns an equally good description of the data, while being simpler in its inherent conception. From a general perspective, we have here reported on a successful attempt to fit an individual (no ensemble averages involved) experimental stochastic trajectory, which displays regular quasi cycles instigated by demographic noise. We hope this triggers renewed interest in data analysis from a fundamentally stochastic angle.

{{\bf Acknowledgements} D.F. thanks L. Chicchi for discussion on a preliminary version of the stochastic model of the Cape Rodney-Okakari Point Marine Reserve.}

\appendix

\section{On the stochastic formulation of population dynamics: a general overview}

The classical approach to biological modeling relies on characterizing  the dynamics of the population under consideration through a system of ordinary differential equations which incorporates the specific interactions at play. As opposed to this formulation, a different level of modeling can be invoked which begins from an individual-based description of the process. This latter description is intrinsically stochastic, as it is subject to finite size fluctuations. This endogenous source of disturbance is also called demographic noise. In the limit of an infinite population, the dynamics becomes deterministic. When the population is instead large but finite, the dynamics is often still strongly stochastic. 

The individual based dynamics can be effectively modeled in terms of simple and rather intuitive chemical-like equations. Random events (e.g. birth, death and predation) occurring at the level of individuals (the microscale) give rise to macroscopic dynamics for large populations of analogous individuals. Denote by $\boldsymbol{n}$ the state vector of the  system under consideration. More specifically, $n_i$, the $i-$th  component of vector $\boldsymbol{n}$, returns the number of elements of  population $i$. Then, we label by $P(\boldsymbol{n},t)$ the probability of finding the system in state $\boldsymbol{n}$ at time $t$. The time evolution of $P(\boldsymbol{n},t)$ is governed by the so called master equation which can be cast in the general form: 

\begin{equation}
\label{eq: ME}
    \frac{\partial P(\boldsymbol{n},t)}{\partial t}= \sum_{\boldsymbol{n'}\ne \boldsymbol{n}} \Bigl[ T(\boldsymbol{n} \vert \boldsymbol{n'}) P(\boldsymbol{n'},t)- T(\boldsymbol{n'} \vert \boldsymbol{n})P(\boldsymbol{n},t) \Bigr]. \    
\end{equation}

This is a balance of opposite contributions: the transitions {\it towards} the reference state (the associated terms with a plus sign), and the transitions {\it from} the reference state (terms with a minus). Indeed, 
$T(\boldsymbol{n'} \vert \boldsymbol{n})$ (resp. $T(\boldsymbol{n} \vert \boldsymbol{n'})$) is the transition rate, the probability per unit of time that a transition takes place from the initial configuration $\boldsymbol{n}$ (resp. $\boldsymbol{n'}$) towards the final state $\boldsymbol{n'}$ (resp. $\boldsymbol{n}$), compatible with the former via the imposed chemical equations. The stochiometric coefficients are stored in a vector ($\boldsymbol{\nu}$) and are used to update the status of the system. 

Master equations can only be solved exactly in very few cases, and therefore different schemes to obtain approximate analytical and numerical solutions have been devised. One such scheme is the linear noise approximation that we have used in the paper. 

Individual realisations of the dynamics can be also obtained via Monte Carlo simulations, which outputs a stochastic time series of the studied model.

\section{The Lotka-Volterra model: microscopic formulation}

For this simplified illustration, we model a population of prey ($X$) and predator ($Y$) individuals in a system of fixed size $N$. We assume that prey can only die because of predation, and that all predation interactions result in the birth of a predator individual. The individual processes governing the dynamics of the system, and their corresponding stoichiometric vectors, are as follows. 

The birth of a prey individual:
\begin{equation}
X \xrightarrow{\beta} X + X ; \quad T_1(n_{X} + 1, n_{Y}\vert \bm n) = \beta \frac{n_X}{N}\frac{(N-n_X-n_Y)}{N} ; \quad  \bm \nu_1 = (1, 0).
\end{equation}

A predator eats a prey, and a new predator is born:
\begin{equation}
X + Y \xrightarrow{\gamma} Y + Y ; \quad T_2(n_{X} -1, n_{Y}+1\vert \bm n) = \gamma \frac{n_X}{N}\frac{n_Y}{N} ; \quad  \bm \nu_2 = (-1, 1).
\end{equation}

A predator dies:
\begin{equation}
Y \xrightarrow{\delta} \varnothing ; \quad T_3(n_{X}, n_{Y}-1\vert \bm n) = \delta \frac{n_Y}{N} ; \quad  \bm \nu_3 = (0, -1).
\end{equation}

Applying the methodology described in Refs.\cite{mckane2014,BLACK2012337}
 gives Eqs. [5] in the paper, where 
we have set $\delta=1$.   

\section{Microscopic interactions for the dynamics of the Cape Rodney-Okakari Point Marine reserve}

We model the interactions between the species in the Cape Rodney-Okakari Point Marine ecological reserve as a set of microscopic reactions that modify the population numbers at the individual level. The different processes taking place in the system, with their corresponding chemical reaction, transition rate and stoichiometric vector are listed in the following.

The mechanism which results in the barnacles to take over the rock can be modeled as:  

\begin{equation}
B_0 + R \xrightarrow{C_{br}} B_0 + B_0 ; \quad T_1(n_{B_0} + 1, n_{B_A}, n_{A_0}, n_M, n_R - 1\vert \bm n) = C_{br} \frac{n_{B_0}}{N}\frac{n_R}{N} ; \quad  \bm \nu_1 = (1, 0, 0, 0, -1).
\end{equation}

Algae colonize the barnacles following two possible pathways. The first read:
  
\begin{equation}
 B_0 + A_0 \xrightarrow{C_{ba}} A_0 + B_A ; \quad  T_2(n_{B_0} - 1, n_{B_A} + 1, n_{A_0}, n_M, n_R\vert \bm n) = C_{ba} \frac{n_{B_0}}{N}\frac{n_{A_0}}{N} ; \quad \bm \nu_2 = (-1, 1, 0, 0, 0).
\end{equation}

while the second can be cast in the form:

\begin{equation}
B_0 + B_A \xrightarrow{C_{ba}} B_A + B_A ; \quad T_3(n_{B_0} - 1, n_{B_A} + 1, n_{A_0}, n_M, n_R\vert \bm n) = C_{ba} \frac{n_{B_0}}{N}\frac{n_{B_A}}{N}  ; \quad \bm \nu_3 = (-1, 1, 0, 0, 0).
\end{equation}
    
As mentioned earlier, a mussel can colonize a barnacle and causes its consequent death. We hence postulate:
    
\begin{equation}
M + B_0 \xrightarrow{C_{m}} M + M ; \quad T_4(n_{B_0} - 1, n_{B_A}, n_{A_0}, n_M + 1, n_R\vert \bm n) = C_{m} \frac{n_{B_0}}{N}\frac{n_{M}}{N} ; \quad \bm \nu_4 = (-1, 0, 0, 1, 0).
\end{equation}

When a barnacle dies, it frees an element of rock:
    
\begin{equation}
B_0 \xrightarrow{m_{b}} R ; \quad T_5(n_{B_0} - 1, n_{B_A}, n_{A_0}, n_M, n_R+1\vert \bm n) = m_{b} \frac{n_{B_0}}{N}; \quad \bm \nu_5 = (-1, 0, 0, 0, 1).
\end{equation}

On the other hand, a mussel can invade a patch where algae and barnacle coexisted:
   
\begin{equation}
 M + B_A \xrightarrow{C_{m}} M + M ; \quad T_6(n_{B_0}, n_{B_A} - 1, n_{A_0}, n_M + 1, n_R\vert \bm n) = C_{m} \frac{n_{M}}{N}\frac{n_{B_A}}{N} ; \quad \bm \nu_6 = (0, -1, 0, 1, 0).
 \end{equation}

Algae present on a barnacle can in turn colonise an adjacent patch of rock. To keep track of this possibility without explicitly introducing the notion of space we set:
    
\begin{equation} 
R + B_A \xrightarrow{C_{ar}} A_0 + B_A; \quad T_7(n_{B_0}, n_{B_A}, n_{A_0}+1, n_M, n_R-1\vert \bm n) = C_{ar} \frac{n_{R}}{N}\frac{n_{B_A}}{N} ; \quad \bm \nu_7 = (0, 0, 1, 0, -1).
\end{equation}

When barnacles covered by the algae colonize a bare rock, the following reaction takes place:
                 
\begin{equation} 
R + B_A \xrightarrow{C_{br}} B_0 + B_A ; \quad T_8(n_{B_0} + 1, n_{B_A}, n_{A_0}, n_M, n_R-1\vert \bm n) = C_{br} \frac{n_{R}}{N}\frac{n_{B_A}}{N} ; \quad \bm \nu_8 = (1, 0, 0, 0, -1).\end{equation}
    
Algae on a barnacles can in turn die:

\begin{equation} 
B_A \xrightarrow{m_{a}} B_0; \quad T_9(n_{B_0} + 1, n_{B_A} - 1, n_{A_0}, n_M, n_R\vert \bm n) = m_{a} \frac{n_{B_A}}{N} ; \quad \bm \nu_9 = (1, -1, 0, 0, 0).
\end{equation} 
    
 Barnacles covered by the algae experience the same fate:
    
 \begin{equation} 
 B_A \xrightarrow{m_{b}} A_0 ; \quad T_{10}(n_{B_0}, n_{B_A} - 1, n_{A_0} + 1, n_M, n_R\vert \bm n) = m_{b} \frac{n_{B_A}}{N} ; \quad \bm \nu_{10} = (0, -1, 1, 0, 0).
  \end{equation} 

Algae anchored to the rock can extend to a neighbor rock patch:
            
\begin{equation} 
 A_0 + R \xrightarrow{C_{ar}} A_0 + A_0; \quad T_{11}(n_{B_0}, n_{B_A}, n_{A_0}+1, n_M, n_R-1\vert \bm n) = C_{ar} \frac{n_{A_0}}{N}\frac{n_{R}}{N}; \quad \bm \nu_{11} = (0, 0, 1, 0, -1).
 \end{equation} 
    
Mussels cause the death of the algae:
                
\begin{equation} 
 A_0 + M \xrightarrow{C_{m}} M + M ; \quad T_{12}(n_{B_0}, n_{B_A}, n_{A_0}-1, n_M+1, n_R\vert \bm n) = C_{m} \frac{n_{A_0}}{N}\frac{n_{M}}{N}; \quad \bm \nu_{12} = (0, 0, -1, 1, 0).
\end{equation} 

When a algae dies rock is freed:

\begin{equation} 
 A_0 \xrightarrow{m_{a}} R; \quad T_{13}(n_{B_0}, n_{B_A}, n_{A_0}-1, n_M, n_R+1\vert \bm n) = m_{a} \frac{n_{A_0}}{N}; \quad \bm \nu_{13} = (0, 0, -1, 0, 1).
 \end{equation} 

Similarly, when a mussel dies an element of rock is freed:

\begin{equation} 
M \xrightarrow{m_{m}} R; \quad T_{14}(n_{B_0}, n_{B_A}, n_{A_0}, n_M-1, n_R+1\vert \bm n) = m_{m} \frac{n_{M}}{N}; \quad \bm \nu_{14} = (0, 0, 0, -1, 1).
\end{equation} 

Algae can also come from the outside, but need bare rocks to colonize:

\begin{equation} 
R \xrightarrow{\alpha_{a}} A_0; \quad T_{15}(n_{B_0}, n_{B_A}, n_{A_0}+1, n_M, n_R-1\vert \bm n) = \alpha_{a} \frac{n_{R}}{N} ; \quad \bm \nu_{15} = (0, 0, 1, 0, -1).
\end{equation} 
    
Analogously, a barnacle can invade the system from the outside provided a rock patch is eventually available:  

\begin{equation} 
R \xrightarrow{\alpha_{b}} B_0; \quad T_{16}(n_{B_0} + 1, n_{B_A}, n_{A_0}, n_M, n_R-1\vert \bm n) = \alpha_{b} \frac{n_{R}}{N}; \quad \bm \nu_{16} = (1, 0, 0, 0, -1).
\end{equation} 

Algae can also come from the outside, but need barnacles to immigrate:

\begin{equation} 
B_0 \xrightarrow{\alpha_{a}} B_A; \quad T_{17}(n_{B_0} - 1, n_{B_A} + 1, n_{A_0}, n_M, n_R\vert \bm n) = \alpha_{a} \frac{n_{B_0}}{N} ; \quad \bm \nu_{17} = (-1, 1, 0, 0, 0).
\end{equation} 
 
Mussels can enter the system, via three alternative pathways listed in the following:

\begin{equation} 
B_0 \xrightarrow{\alpha_{m}} M ; \quad T_{18}(n_{B_0} - 1, n_{B_A}, n_{A_0}, n_M+1, n_R\vert \bm n) = \alpha_{m} \frac{n_{B_0}}{N}; \quad \bm \nu_{18} = (-1, 0, 0, 1, 0).
\end{equation}

\begin{equation} 
 B_A \xrightarrow{\alpha_{m}} M ; \quad T_{19}(n_{B_0}, n_{B_A} - 1, n_{A_0}, n_M+1, n_R\vert \bm n) = \alpha_{m} \frac{n_{B_A}}{N} ; \quad \bm \nu_{19} = (0, -1, 0, 1, 0).
 \end{equation}

\begin{equation} 
 A_0 \xrightarrow{\alpha_{m}} M; \quad T_{20}(n_{B_0}, n_{B_A}, n_{A_0} - 1, n_M+1, n_R\vert \bm n) = \alpha_{m} \frac{n_{A_0}}{N} ; \quad \bm \nu_{20} = (0, 0, -1, 1, 0).
 \end{equation}

Applying the methodology described in Refs.\cite{mckane2014,BLACK2012337}
 gives Eqs. [10] and [11] in the paper, where 
we have set $\delta=1$.

\section{Data normalization}

Before carrying out the estimation and reconstruction for the real data, we need to preprocess it. The populations are recorded as percentage of coverage at different dates spanning ca.~20 years. The $B=b_0+b_A$, $A=b_A+a_0$ and $r$ populations are the average of 10 `observation patches', while the data for $m$ correspond to the average of 5 of such patches. We use the interpolated time series from~\cite{beninca2015}, which has values for all populations at 30-day intervals. The relationship between the data and the model is given by
\begin{align*}
B &= (b_0 + b_A)\, \times 100\%\\
A &= (a_0 + b_A)\, \times 100\%\\
M &= m\, \times 100\%\\
R &= r\, \times 100\%.
\end{align*}
Thus, every data point should satisfy $B+A+M+R = 100(1+b_A)$. For a data point to be correctly normalized, we should have:
\begin{enumerate}
\item $B + A + M + R \geq 100$
\item $\max (B,A) + M + R \leq 100$.
\end{enumerate}
Of the 251 total entries in the interpolated dataset, we find that more than half of them are incorrectly normalized. The situation is fine for the first of the conditions above: the latest data point with this problem comes just after Dec 1995---when a change in the procedure used to sample the population numbers was introduced~\cite{beninca2015}---and the sum of the variables amounts to almost 100\%. After Jan 1996, there are no errors of this type. If we focus only on dates after Dec 1995, the data points violating the second condition are also just above 100\%, except for a few showing extreme values. Since the parameter estimation method we are using requires that we have information for the same variables at every time point, but not necessarily that the time points are equidistant, we could simply remove data points that surpass a certain threshold; here, we have chosen the threshold to be 105\%.

The full normalization procedure we have applied is as follows:
\begin{enumerate}
\item For correctly normalized data we take $b_A = (B+A+M+R)/100 - 1$. Then $b_0 = B/100 - b_A$ and $a_0 = A/100 - b_A$.
\item For points violating the first condition, the ``excess'' over 100 is negative, which would result in $b_A < 0$. We truncate the values to $b_A = 0$ and then fill the remaining `unoccupied' space with rock by forcing $r = 1 - (B + A + M)/100$.
\item For points violating the second condition, there are two cases:
	\begin{itemize}
	\item If $\max (B,A) + M > 100$, we set $r = 0$ and force $m = 1 - \max(B,A)/100$.
	\item If not, we force $r = 1 - [\max (B,A) + M]/100$.
	\end{itemize}
    After this, we assume that whichever population is smaller between $B$ and $A$ is fully concentrated in $b_A$; that is, if $B>A$ (resp. $A>B$), then $b_A = A/100$, $b_0 = (B - A)/100$ and $a_0 = 0$ (resp. $b_A = B/100$, $a_0 = (A - B)/100$ and $b_0 = 0$).
\end{enumerate}

\begin{figure*}
    \centering
   \includegraphics[width=0.8\textwidth]{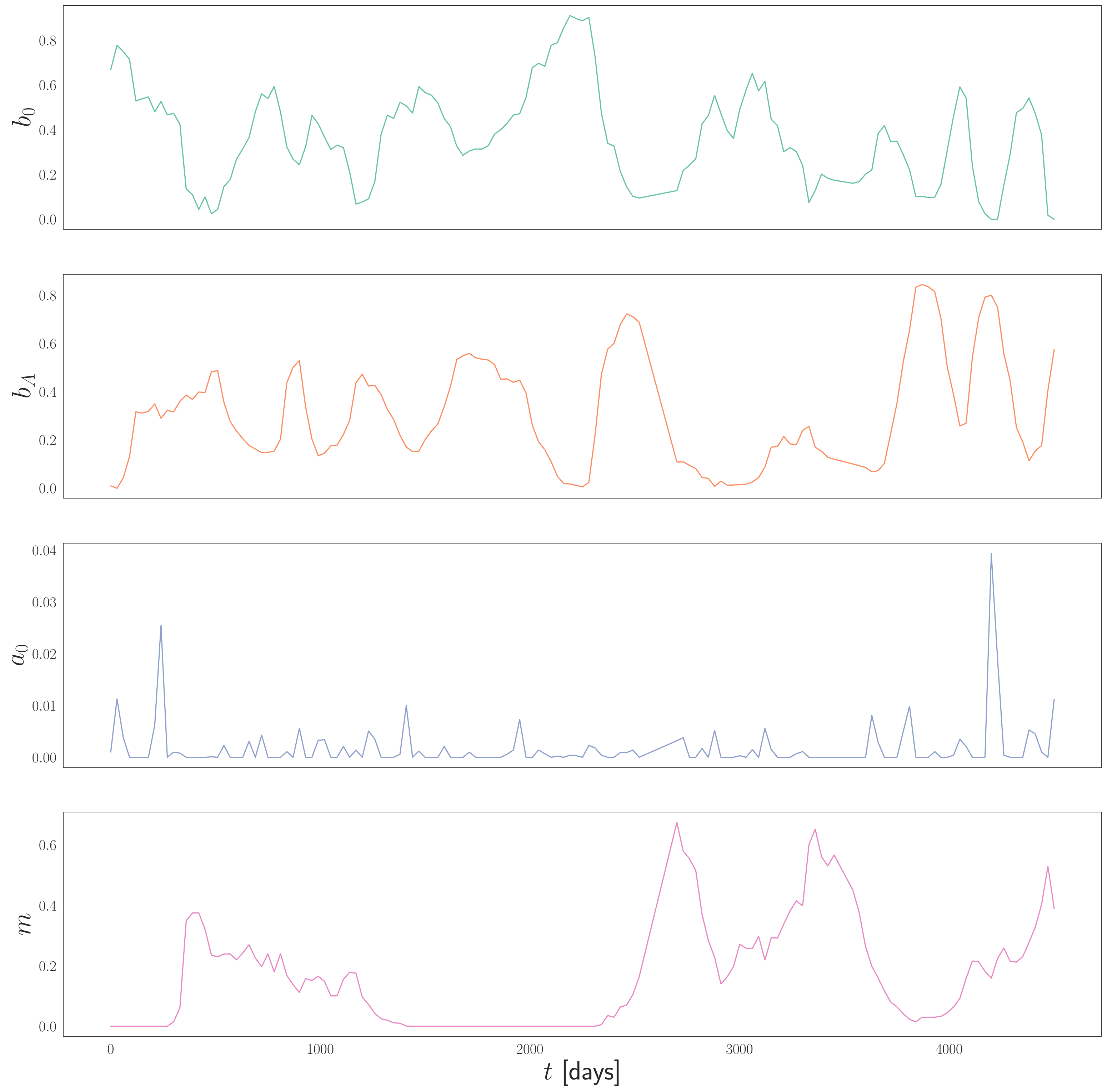}
    \caption{Processed and normalized data from~\cite{beninca2015}.}
    \label{fig:data}
\end{figure*}

The resulting dataset contains 144 entries, and the trajectories are shown in \cref{fig:data}.

%%% Each figure should be on its own page
\begin{figure*}
    \centering
   \includegraphics[width=0.33\textwidth]{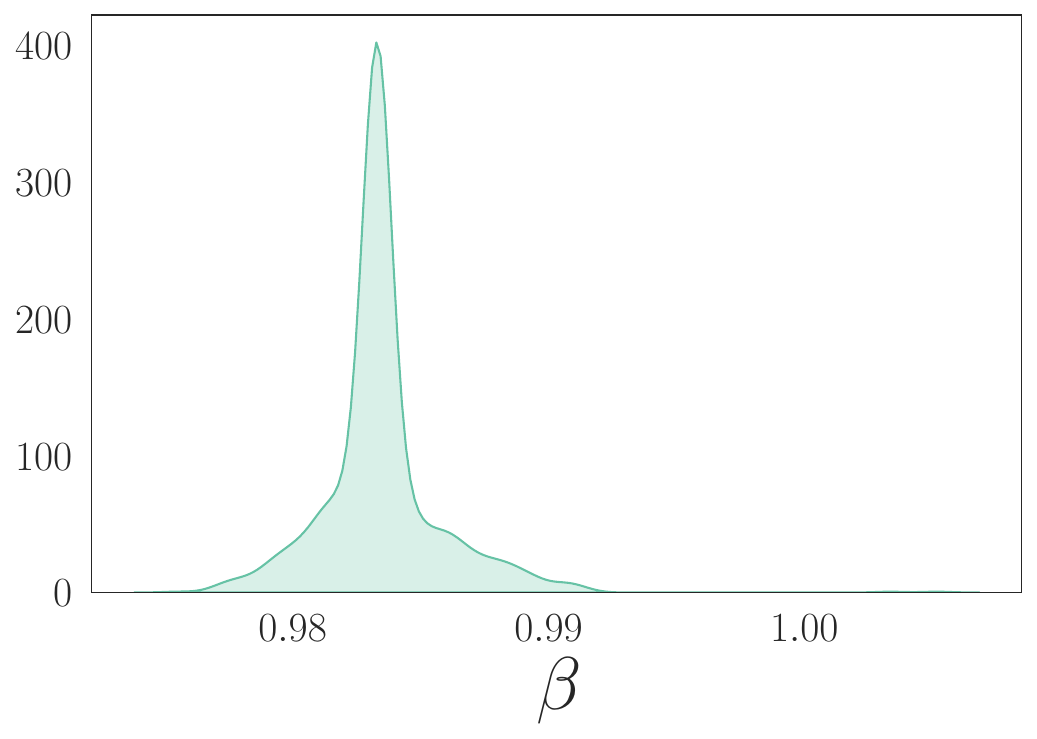}\hfill \includegraphics[width=0.33\textwidth]{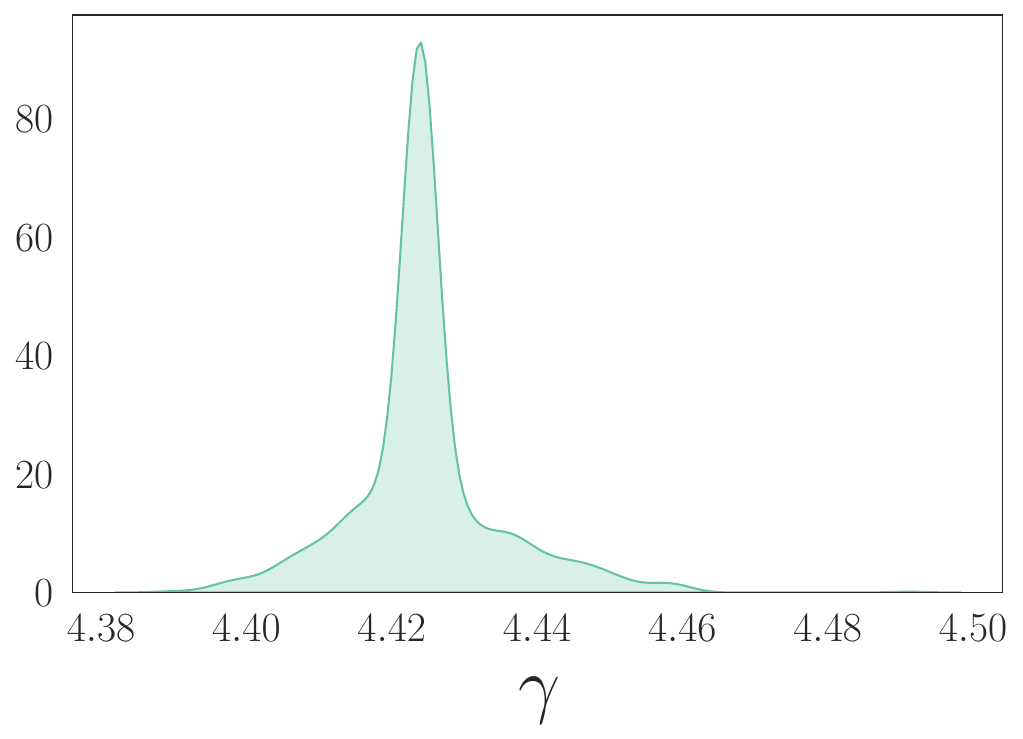}\hfill \includegraphics[width=0.33\textwidth]{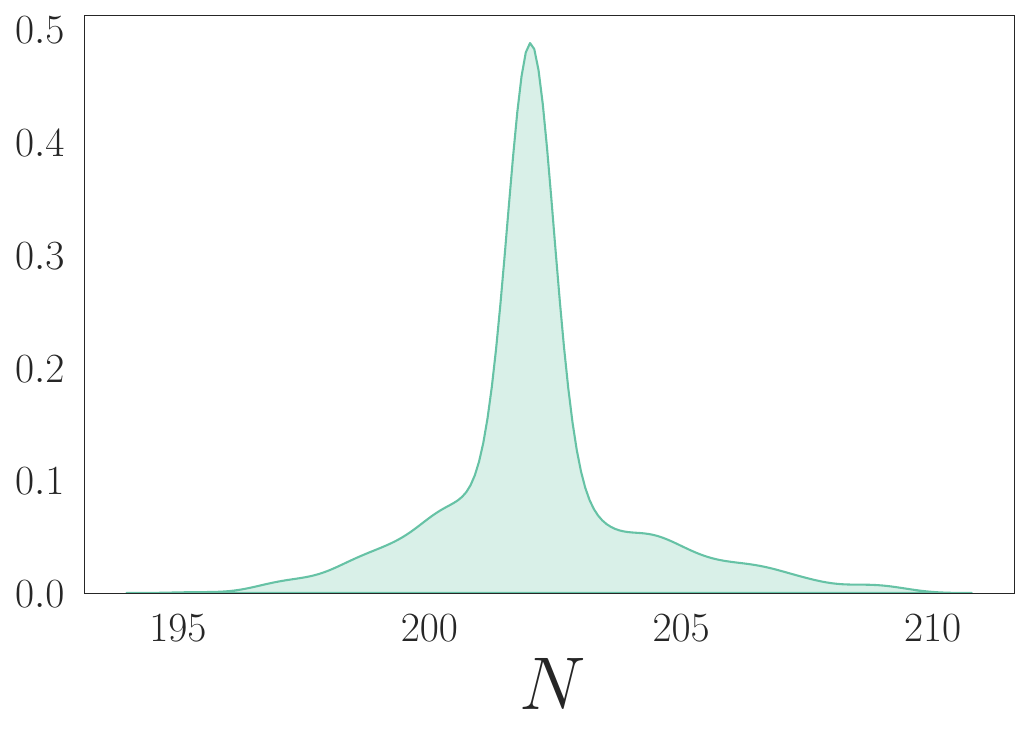}\\
   \includegraphics[width=0.33\textwidth]{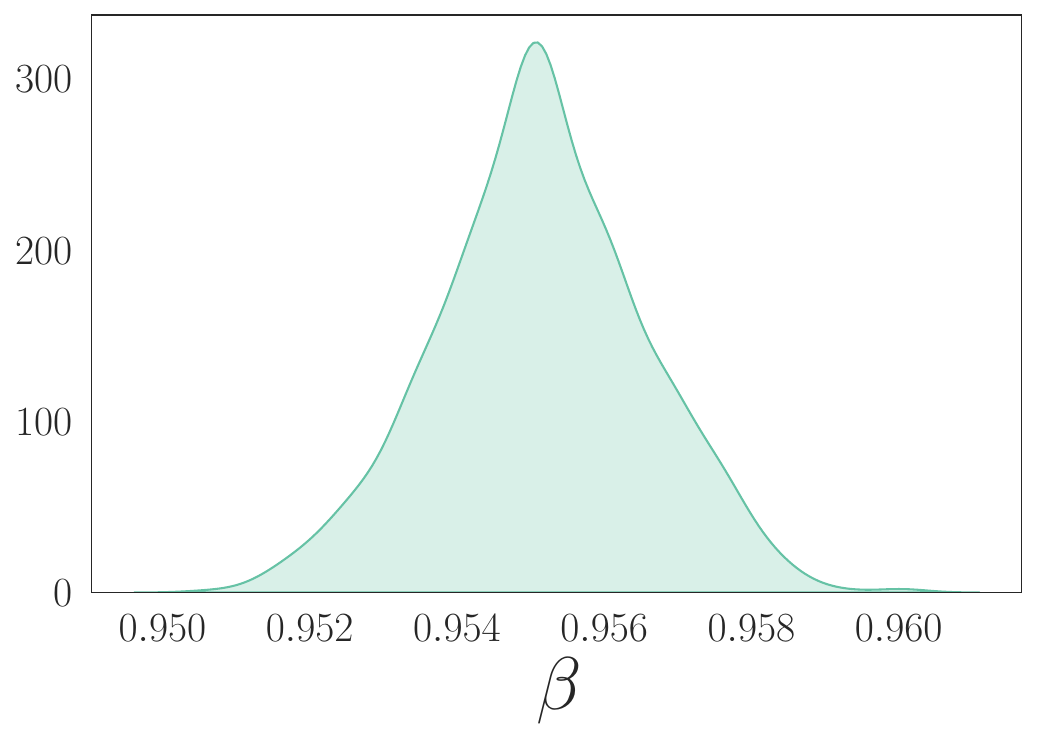}\hfill \includegraphics[width=0.33\textwidth]{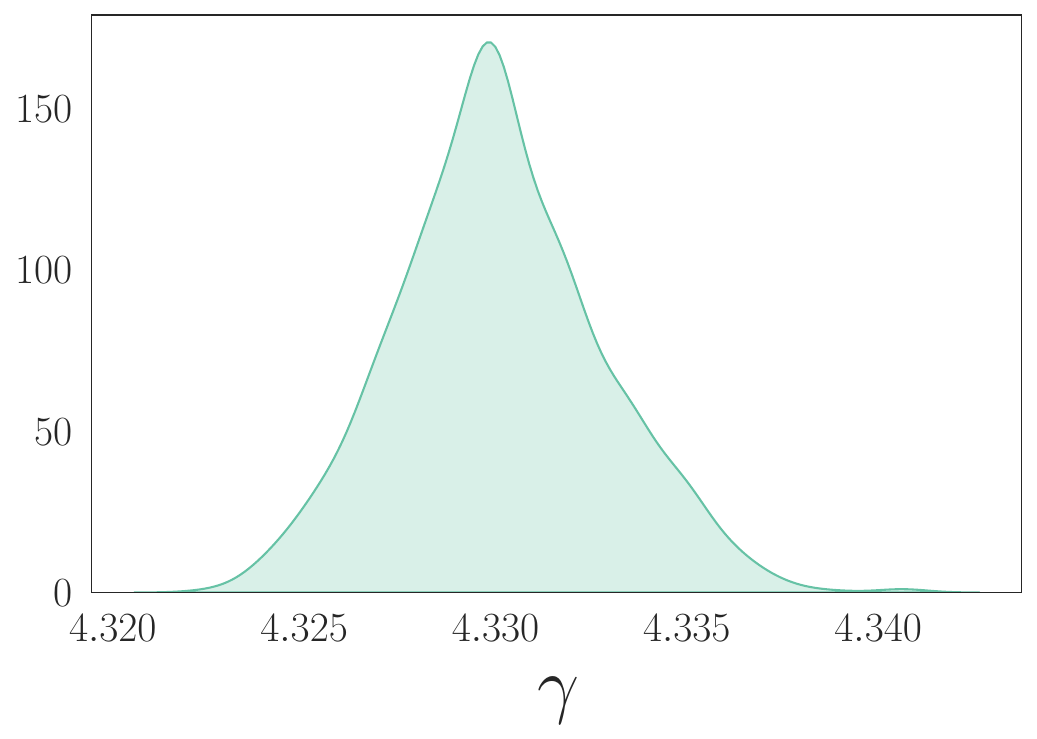}\hfill \includegraphics[width=0.33\textwidth]{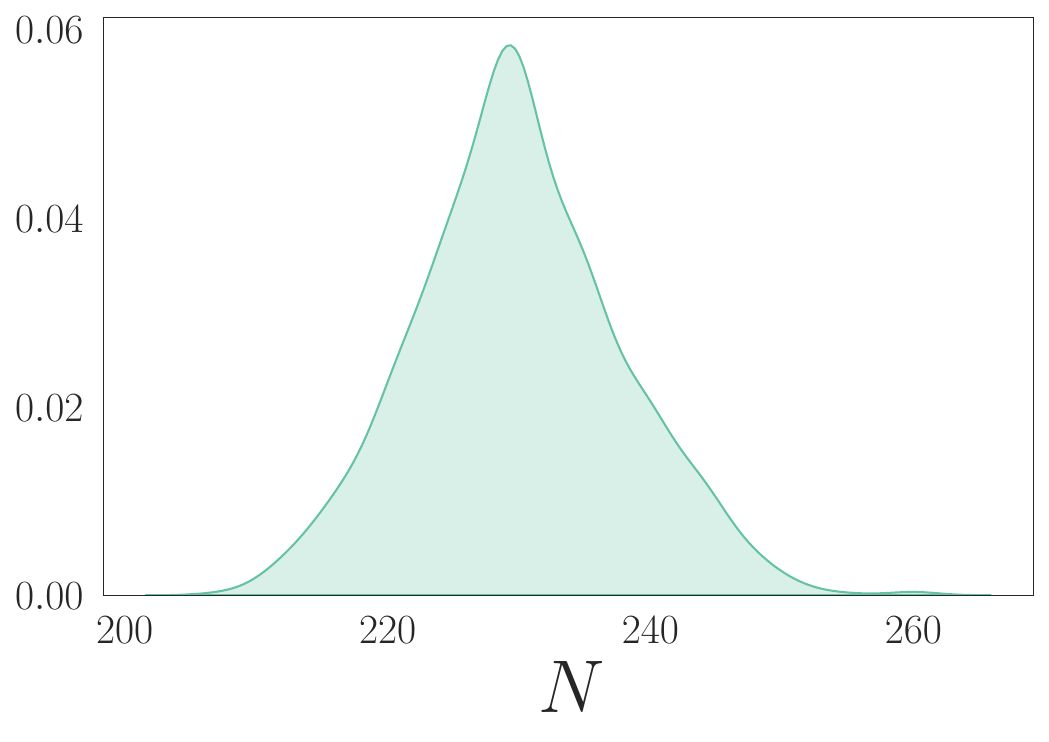}
    \caption{Distribution of the parameters obtained for the data from the Lotka--Volterra model, after estimating it 1200 times with a sampling interval of 10 timesteps (top) and 200 timesteps (bottom). Nominal values: $\beta=1$, $\gamma=4.5$, and $N=200$.}
    \label{fig:toy_estimation}
\end{figure*}

\begin{figure*}
    \centering
   \includegraphics[width=0.9\textwidth]{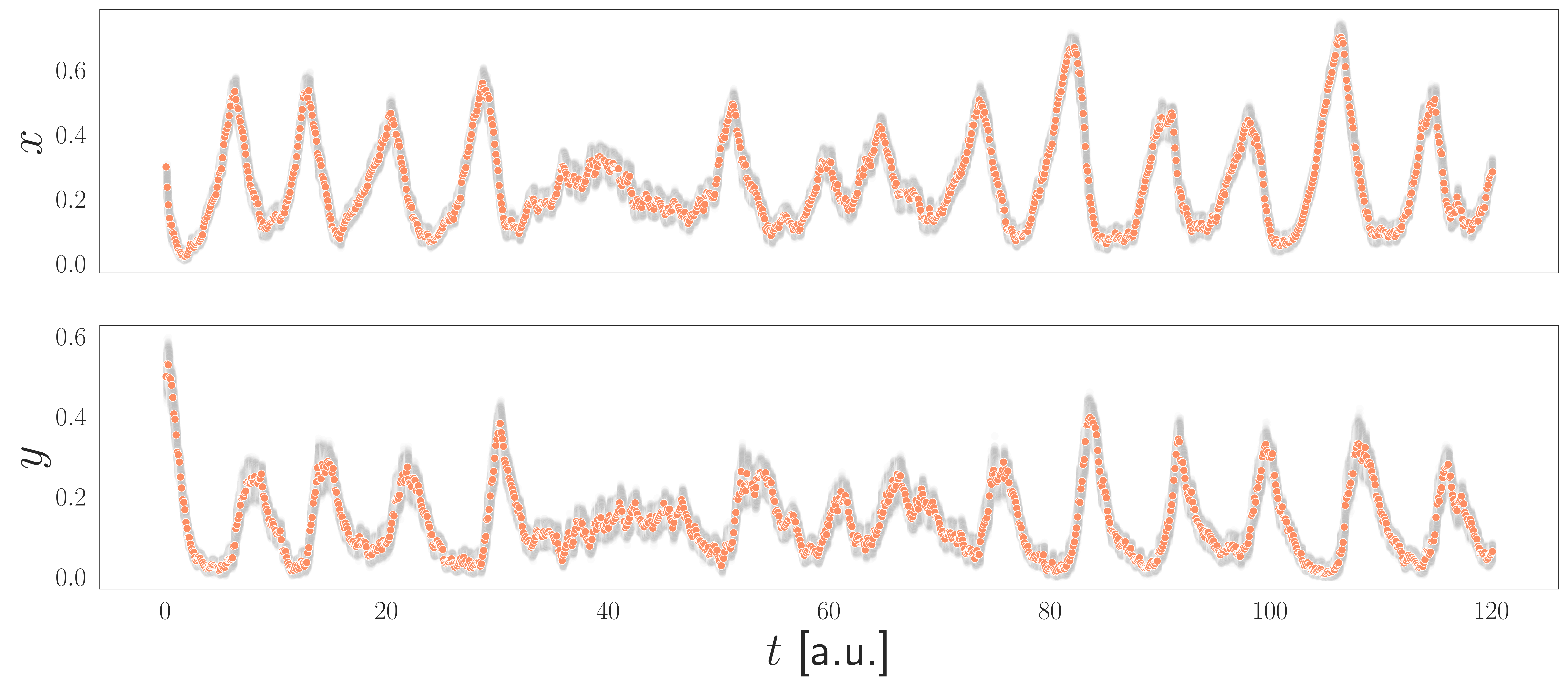}\\
   \includegraphics[width=0.9\textwidth]{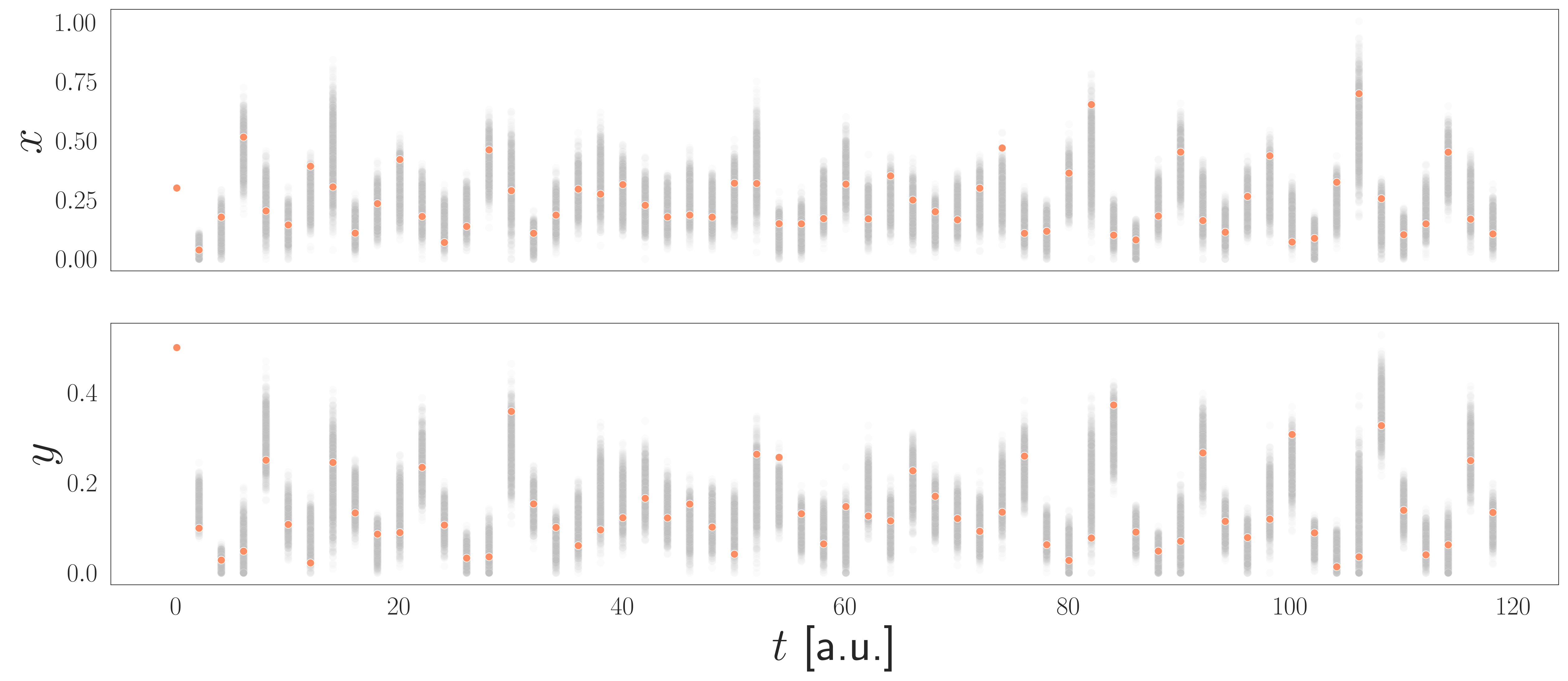}
    \caption{Reconstruction of the Lotka--Volterra time series, using the parameters obtained from the estimation with a sampling interval of 10 timesteps (top) and 200 timesteps (bottom), using 1500 points for the cloud of fluctuations.}
    \label{fig:toy_reconstruction}
\end{figure*}

\bibliography{01-Ref}% Produces the bibliography via BibTeX.

\end{document}